# The Microquasar XTE J1807-294 – Mass Evaluation by Means of the Relativistic Precession Model

Radostina P. Tasheva[1,a)] and Ivan Zh. Stefanov[1, b)]

[1]Department of Applied Physics, Technical University of Sofia, 8 St. Kliment Ohridski Blvd., BG-1000

a)Corresponding author: rpt@tu–sofia.bg
b)izhivkov@tu-sofia.bg

**Abstract.** The high frequency quasiperiodic oscillations (QPOs) in the X-ray spectra of the millisecond pulsar XTE J1807-294 are under consideration. By application of the relativistic precession (RP) model an attempt is made to approximate the observed frequencies as well as to assess the mass and the angular momentum of the object. The obtained mass is too high for a neutron star.

## INTRODUCTION

In the eighties physical society was introduced to an interesting phenomena taking place in both neutron stars (NS) and black holes candidates (BH) existing in low mass binary systems. In their low frequent (LF, 1 – 100 Hz) X–ray spectra appeared nearly periodical spikes called quasiperiodic oscillations (QPOs, see Ref.[1]). About 20 years later the data from the Rossi X-ray Timing Explorer (RXTE) showed QPOs in the kHz range (HF) – from 0.2 to 1.25 kHz. Persistent repetition of both of these oscillations provide the scientists with the opportunity to study phenomena that happen in the innermost part of accretion disks surrounding the central objects where they are believed to originate i.e. in the conditions of extremely strong gravitational fields.

The behavior of the two QPO groups has proved to be completely different – while the LF QPOs are strong, persistent and tend to drift in frequency HF QPOs are transient and weak but do not shift their frequencies significantly. Following these data a hypothesis can be suggested - they are probably created in different parts of the accretion disk.

For some of the low mass X-ray binaries containing NS or BH the lower (L) and the upper (U) HF QPOs exist in pairs. As Kotrlova et al. [2] has demonstrated the simultaneous appearance of L and U HF QPOs allows the mass-angular momentum relation (a-M) to be determined. If for example the mass is obtained using photometric data then angular momentum can be evaluated. In some cases twin kHz QPO are observed in more than one pair. These results allow to construct different mass-angular momentum relations. The extent of their agreement can be used as a testing ground for the model applied to explain the objects` HF QPOs.

The discovery of twin kHz QPOs in the X ray flux of accretion millisecond pulsar XTE J1807-294 is firstly reported in 2005 by Linares et al. [3]. They observed eight different pairs of simultaneous kHz QPOs with ratio $\nu_U/\nu_L$ varying from 3:2 to approximately 3:1. In the present paper we aim by applying the relativistic precession (RP) model to this object and solving the correspondent analytic mass-angular momentum equation to estimate the optimal mass $M_{opt}$ and optimal angular momentum $a_{opt}$ of the object according to the $\chi^2$ test. The comparison between the so obtained values and the results received using other methods (photometrical, spectral etc.) can serve as an assessment of the validity of the model we apply.

The paper is organized as follows. The observations of the twin kHz QPOs in XTE J1807-294 are briefly presented in Section 2. Section 3 deals with the essence of the RP model. Section 4 describes the application of the

$\chi^2$ test to the object. Section 5 discusses the agreement between the observational frequencies and results of the application of the RP model. The last section is the conclusion.

In this paper the masses are given in solar masses, the radii x are scaled with the gravitational radius $r_g \equiv GM/c^2$, the specific angular momentum is $a \equiv cJ/GM^2$ and is accepted that $G = 1 = c$ where $G$ is the universal gravitational constant and $c$ is the speed of light.

## OBSERVATIONAL DATA

XTE J1807–294 is the fourth-discovered accreting millisecond pulsar ( after SAX J1808.4–3658 , XTE J1751–305, and XTE J0929–314). The first detection of the source has been on 13th of February, 2003, during the periodic scans of the Galactic bulge region by means of RXTE. Linares et al. [3] reported that using the Proportional Counter Array (PCA), an instrument on board of RXTE on 21st of February 2003 190.6 Hz pulsations has been discovered thus confirming that the source is an accreting millisecond pulsar (see Ref.[4]).

The data this investigation consists of are taken from 27th of February to 16th of March 2003. In order to fit the power spectrum of each of the eight groups available a multi-Lorentzian function is implemented. Every multi-Lorenzian is assembled by adding several Lorentzians. Each one of the Lorentzians corresponds to a recognizable component in the power density spectrum (PDS).

The eight datasets and the best-fit frequencies of the QPOs are listed in Table 1.The set A according to Linares et al. (2005) do not to show simultaneous kHz QPOs and has to be treated with caution.

**TABLE 1.** Twin kHz QPO frequencies with their uncertainties

| Group | $v_L$(Hz) | $v_U$(Hz) |
|---|---|---|
| A | 106.0±23.0 | 337.0±10.0 |
| B | 163.0±23.0 | 354.0±4.0 |
| C | 191.0±9.0 | 375.0±2.0 |
| D | 202.0±11.0 | 395.0±3.0 |
| E | 238.0±25.0 | 449.0±9.0 |
| F | 259.0±16.0 | 465.0±2.0 |
| G | 273.0±19.0 | 492.0±5.0 |
| H | 370.0±18 | 565.0±5.0 |

## THE RELATIVISTIC PRECESSION MODEL

Neutron stars as well as black holes are final stages of evolution of stars with masses bigger than solar and much denser. If such an object exists in a binary system then accretion disc forms around it. The matter rotating around the central object and falling toward it consists of charged particles attracted by the strong gravitational field.

Let a test particle is assumed orbiting along circular geodesics in the innermost part of the accretion disc i.e. close to the innermost stable circular orbit (ISCO). The values of the high frequencies are close to those of the fundamental frequencies of motion – orbital frequency $v_\phi$, radial $v_r$ and vertical $v_\theta$ of the test particle. This is the reason why the models constructed to explain the creation of HF QPO often exploit simple combinations of fundamental frequencies of particle motion.

The relativistic precession model proposed by Stella and Vietri [5] is the first model that relates the frequency of the LF QPOs, namely $v_{LF}$ to the to frame dragging off equatorial orbits and Lense-Thirring precession. The frame dragging is one of the relativistic effects of the "gravitomagnetic" field associated with rotating bodies. It coerces a test particle moving in a non-equatorial plane to start precession around the angular momentum axis of the rotating object. The value for the low frequency is $v_{LF}=v_{LT} = |v_\phi - v_\theta|$. The RP model accepts that all three QPOs – the low LF and the both high – lower $v_L$ and upper $v_U$ are created at the same orbit.

The higher frequency $v_U$ is a direct result of modulation of the X-ray flux by the orbital frequency and $v_U = v_\phi$ The lower frequency $v_L$ is ascribed to the periastron precession of the relativistic orbit planes i.e $v_{per} = v_\phi - v_r$.
The correlation between $v_{LF}$ and $v_L$ that has been observed for a large sample of neutron stars and black hole candidates has found its satisfactory explanation through application of the RP model (see Ref.[6]).

# X² TEST APPLICATION

The RP model is expected to explain successfully the observational frequencies of the QPOs. It also provides a mass-angular momentum relation. The information about the main parameters of the central object – its mass and angular momentum can be retrieved (without reference to external sources) by checking the datasets for consistency. The $\chi^2$ test can be implemented in order to show if the chosen model is applicable for investigated object. As it concerns the reliability of the so obtained results comparison with photometric or spectral data will show where our results are in conflict with other investigations. The object selected by us – XTE J1807-294 shows commensurability between the lower $\nu_L$ and the higher $\nu_U$ frequency that varies from 3:1 to 3:2. Generally we can express one of the frequencies, for example $\nu_L$ as a function of value $\nu_U$ or vice versa in the following way:

$$\nu_L = f(\nu_U) = \nu_L^{obs}, \qquad \nu_U = f(\nu_L) = \nu_U^{obs}. \tag{1}$$

Both frequencies depend on the free parameters - the mass M and the specific angular momentum a of the neutron star. We use the experimental values $\{\nu_{L,i}^{obs}, \nu_{U,i}^{obs}\}$ of the frequencies, i=1, 2, 3…N in order to find the optimal values $M_{opt}$ and $a_{opt}$ for the free parameters in the chosen model. If the $\chi^2$ test is applied to either of the above equations the solutions for $M_{opt}$ and $a_{opt}$ will allow us to draw the line of best fit for the both relations. This will be done by minimization of the functions:

$$\chi_L^2(a,M) = \sum_{i=1}^{N} \frac{\left(\nu_L(a,M,\nu_U) - \nu_{L,i}^{obs}\right)^2}{\sigma_{L,i}^2}, \qquad \chi_U^2(a,M) = \sum_{i=1}^{N} \frac{\left(\nu_U(a,M,\nu_L) - \nu_{U,i}^{obs}\right)^2}{\sigma_{U,i}^2}. \tag{2}$$

In the first of Eq. (2) $\nu_L$ is the dependent variable and $\nu_U$ – the independent one, in the next equation – vice versa. If the expression contains N=8 pairs of frequencies and M=2 free parameters (a, M) i.e. N-M=6 degrees of freedom i.e. the acceptable values corresponding to 90% confidence level for $\chi^2$ are $0 \leq \chi^2 \leq 10.6$. For N=7 pairs i.e. 5 degrees of freedom $0 \leq \chi^2 \leq 9.24$.

According to the RP model the both high frequencies originate at the same radius $x_L = x_U$ and the observational frequencies depend in the same way on x. Then Eq. (1) can be written in the following form:

$$\nu_L(\nu_U^{obs}) = f\left(x_U(\nu_U^{obs})\right), \qquad \nu_U(\nu_L^{obs}) = f\left(x_L(\nu_L^{obs})\right).$$

A substantial problem for the application of the $\chi^2$ causes the fact that only the standard error of the dependent variable is considered during the calculation while the standard error of the independent variable is neglected.

# RESULTS AND DISSCUSION

The $\chi^2$ test is applied firstly to eight groups of data according to Linares et.al. (2005) and afterward the test is repeated with seven groups of data. The results of the both cases define the lines of best fit – firstly for $\nu_L = f(\nu_U)$ and then for $\nu_U = f(\nu_L)$ in order to get more reliable and independent of the intrinsic uncertainties results. Dataset A is excluded in the second test because according to the same paper the twin high frequencies in set A might not arise simultaneously.

The dependence of the lower frequency $\nu_L$ as a function of the upper frequency $\nu_U$ is given in the Figure 1 The received value for $a_{opt}$ in the cases where the all datasets are included and when the A data set is excluded are $a_{opt}=0.94\pm0.01$ and $a_{opt}=0.97\pm0.01$ respectively. The mass estimates corresponding to both cases are $M_{opt}=(10.0\pm0.4)$ $M_\odot$ and $M_{opt}=(10.9\pm0.4)$ $M_\odot$ respectively. The optimal mass $M_{opt}$ and angular momentum $a_{opt}$ when eight groups of data are used are defined by $\chi^2_{min}=4.2$. When the number of datasets is seven, significant reduction of the $\chi^2$ is achieved - $\chi^2_{min}=1.5$ i.e. dataset A could be anomalous one. Both results are more than agreeable – the reference values are $\chi^2 \leq 10.6$ and $\chi^2 \leq 9.24$ respectively.

The dependence of the upper frequency $\nu_U$ as a function of the upper frequency $\nu_L$ is given in the Figure 2. The received values for $a_{opt}$ in the cases where the all datasets are included and when the A data set is excluded are $a_{opt}=0.95\pm0.01$ and $a_{opt}=0.98\pm0.01$ respectively. The mass estimates corresponding to both cases are $M_{opt}=(10.4\pm0.1)$ $M_\odot$ and $M_{opt}=(11.2\pm0.1)$ $M_\odot$ respectively.

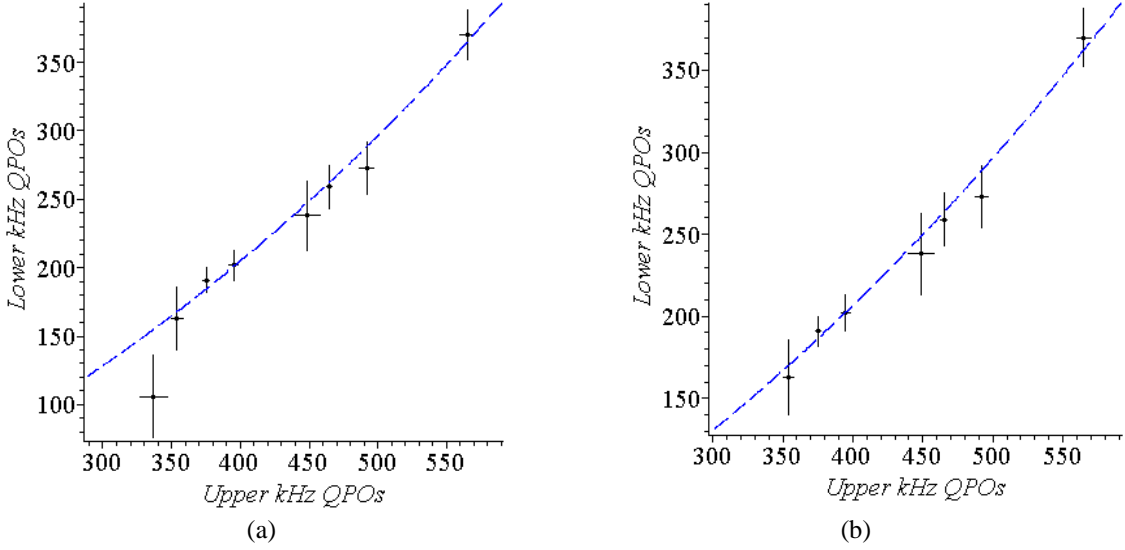

**FIGURE 1.** The dependence of lower $\nu_L$ versus upper $\nu_U$ frequency according to RP model. The representation is a dashed line, . The positions for the experimental frequencies coming from the different groups are given with their uncertainties. (a) – from A to H; (b) – from B to H.

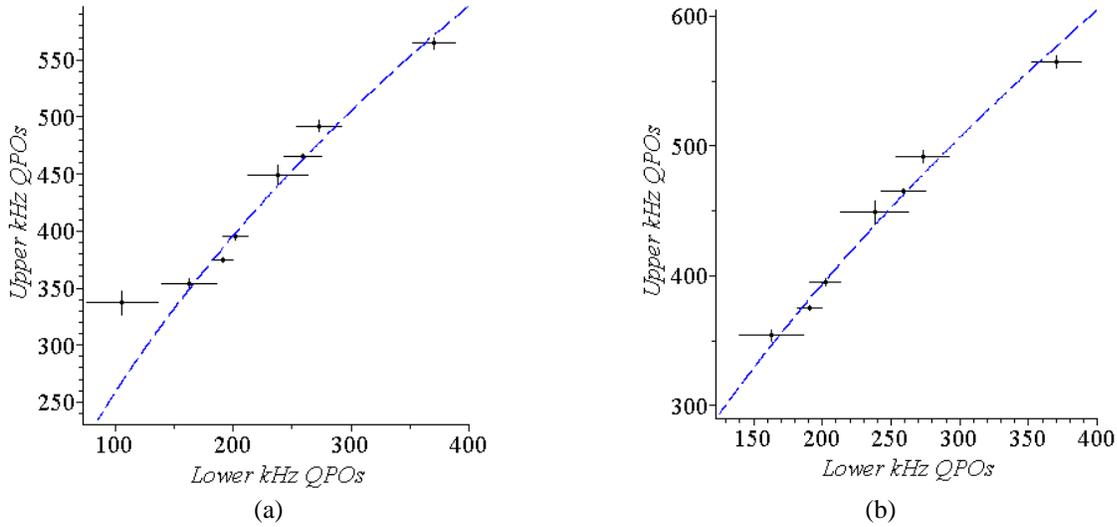

**FIGURE 2.** The dependence of upper $\nu_U$ versus lower $\nu_L$ frequency according to RP model. The representation is a dashed line . The positions for the experimental frequencies coming from the different groups are given with their uncertainties. (a) – from A to H; (b) – from B to H.

The minimum value for $\chi^2_{min}$=85.57 in the first case and is again significant reduced to $\chi^2_{min}$=33.33 in the second case. The line that depicts the expression $\nu_U=f(\nu_L)$ is determined with a much bigger $\chi^2_{min}$ than is calculated for the expression $\nu_L=f(\nu_U)$. The discrepancy may result from the fact that when the calculations are made the standard error for the independent variable is neglected. In the second case $\nu_L$ happens to be independent variable. If we accept coarsely that if the correspondent frequencies are independent variables their absolute uncertainties are respectively $\Delta\nu_{L,ave}= (\Delta\nu_{L,max} + \Delta\nu_{L,min})/2=17.0$ and $\Delta\nu_{U,ave}== (\Delta\nu_{U,max} + \Delta\nu_{U,min})/2=6.0$ then $\Delta\nu_{L,ave} / \Delta\nu_{U,ave} \approx 3$ which may cause a significant difference between $\chi^2$ results in both cases.

Our calculations suggest that the mass of XTE J1807-294 is too big for a neutron star. One possible explanation seems to be the fact that during the retrieval of the data through the $\chi^2$ the uncertainties of the independent variable have to be neglected. Another source of error is non-simultaneous creation of some of the twin QPO.

The applicability of the RP model to this object may also be carefully reconsidered. The orbital motion frequencies near the inner edge of an accretion disc are highly susceptible to even the smallest radial perturbations which can have a strong effect over radial oscillation. Therefore some non-geodesic amendments can be made in order to adjust the models to the observational data. (see Ref.[7]).

The resonant switch model by Stuchlık et al. (see Ref.[8]) for example suggests that more than one model can be applied if the frequency ratio $\nu_U / \nu_L$ changes. In our case $\nu_U / \nu_L > 2$ for A and B datasets and $\nu_U / \nu_L < 2$ for groups C-H i.e. the next step in further investigation could be to consider a suitable resonant switch model.

## CONCLUSION

We applied the RP model to the low mass X-ray binary XTE J1807-294. Using the data of Linares et al. (2005) for the twin QPOs we constructed graphs that represent dependence between the lower $\nu_L$ and the upper $\nu_U$ HF QPOs. Firstly $\nu_U$ is the independent variable and $\nu_L$ is the dependent i.e $\nu_L = f(\nu_U)$ and secondly – vice versa i.e. $\nu_U = f(\nu_L)$. For both cases we implemented the $\chi^2$ test in order to obtain the optimal mass $M_{opt}$ and optimal angular momentum $a_{opt}$ - the ones for which $\chi^2 = \chi^2_{min}$. The results from the $\chi^2$ are more than agreeable for the first case - $\chi^2_{min} = 1.5$ and worse for the second - $\chi^2_{min} = 33.33$ (reference value $\chi^2 \leq 9.24$).

The obtained masses - $M_{opt} = (10.9 \pm 0.4)$ $M_\odot$ in the first case and $M_{opt} = (11.2 \pm 0.1)$ $M_\odot$ in second (for seven datasets) are though too big for neutron stars and the reasons could be multiple - neglected uncertainties of the independent variable, non-simultaneous HF QPOs, inappropriate choice of model. For further investigations an amendment that includes the neglected uncertainties can be used. The switch resonant model also seems to be a viable option to have in mind.

## ACKNOWLEDGMENTS

This research is partially supported by the Bulgarian National Science Fund under Grant No N 12/11 from 20 December 2017.